\documentclass[aps,eqsecnum,preprint,floats,epsf,epsfig,nofootinbib]{revtex4}
\usepackage{epsfig}
\usepackage{graphicx}

\begin{document}
\def\be{\begin{eqnarray}}
\def\en{\end{eqnarray}}
\def\non{\nonumber}
\def\la{\langle}
\def\ra{\rangle}
\def\A{{\cal A}}
\def\B{{\cal B}}
\def\c{{\cal C}}
\def\d{{\cal D}}
\def\e{{\cal E}}
\def\p{{\cal P}}
\def\t{{\cal T}}
\def\nc{N_c^{\rm eff}}
\def\CP{{\it CP}~}
\def\CPP{{\it CP}}
\def\acp{{\cal A}_{C\!P}}
\def\vp{\varepsilon}
\def\drho{\bar\rho}
\def\deta{\bar\eta}
\def\vma{{_{V-A}}}
\def\vpa{{_{V+A}}}
\def\J{{J/\psi}}
\def\ov{\overline}
\def\Lqcd{{\Lambda_{\rm QCD}}}
\def\pr{{ Phys. Rev.}~}
\def\prl{{ Phys. Rev. Lett.}~}
\def\pl{{ Phys. Lett.}~}
\def\np{{ Nucl. Phys.}~}
\def\zp{{ Z. Phys.}~}
\def\lsim{ {\ \lower-1.2pt\vbox{\hbox{\rlap{$<$}\lower5pt\vbox{\hbox{$\sim$}
}}}\ } }
\def\gsim{ {\ \lower-1.2pt\vbox{\hbox{\rlap{$>$}\lower5pt\vbox{\hbox{$\sim$}
}}}\ } }

\font\el=cmbx10 scaled \magstep2{\obeylines \hfill May, 2020}

\vskip 1.0 cm

\centerline{\large\bf {\it CP} Violation
in $B^\pm\to \rho^0\pi^\pm$ and $B^\pm\to \sigma\pi^\pm$ Decays}
\bigskip
\centerline{\bf Hai-Yang Cheng}
\medskip
\centerline{Institute of Physics, Academia Sinica}
\centerline{Taipei, Taiwan 115, Republic of China}
\medskip
\bigskip

\bigskip
\bigskip
\centerline{\bf Abstract}
\bigskip

\small
The decay amplitude of $B^+\to \pi^+\pi^-\pi^+$ in the Dalitz plot has been analyzed by the LHCb using three different approaches for the $S$-wave component. It was found that the mode with $\sigma$ (or $f_0(500)$) exhibited a \CP asymmetry of 15\% in the isobar model, whereas the $f_2(1270)$ mode had a 40\% asymmetry. On the contrary,
\CP asymmetry for the dominant quasi-two-body decay $B^-\to\rho^0\pi^-$ was found to  be consistent with zero in all three approaches, while all the existing theoretical
predictions  lead to a negative \CP asymmetry ranging from $-7\%$ to $-45\%$. We show that the nearly vanishing \CP violation in $B^-\to\rho^0\pi^-$ is understandable in the framework of QCD factorization (QCDF). It arises from the $1/m_b$ power corrections to the penguin amplitudes due to penguin annihilations and to the color-suppressed tree amplitude due to hard spectator interactions. Penguin annihilation and hard spectator interactions contribute destructively to $\acp(B^-\to\rho^0\pi^-)$ to render it consistent with zero. The branching fraction and \CP asymmetry in $B^-\to\sigma/f_0(500)\pi^-$ are investigated in QCDF with results in agreement with experiment.

\pagebreak

\section{Introduction}

In 2013 and 2014 LHCb has measured direct \CP violation in charmless three-body decays of $B$ mesons
\cite{LHCb:Kppippim,LHCb:pippippim,LHCb:2014} and found evidence of inclusive integrated \CP asymmetries in $B^+\to\pi^+\pi^+\pi^-$, $K^+K^+K^-$, $K^+K^-\pi^+$ and a 2.8$\sigma$ signal of
\CP violation in $B^+\to K^+\pi^+\pi^-$.  Besides the integrated \CP asymmetry,
LHCb has also observed large asymmetries in localized regions of phase space, such as the low invariant mass region devoid of most of known resonances and the rescattering regions of of $m_{\pi^+\pi^-}$ or $m_{K^+K^-}$ between 1.0 and 1.5 GeV.

Recently LHCb has analyzed the decay amplitudes of $B^+\to \pi^+\pi^-\pi^+$ in the Dalitz plot \cite{Aaij:3pi_1,Aaij:3pi_2}.
In the LHCb analysis, the $S$-wave component of $B^-\to \pi^+\pi^-\pi^-$ was studied using three different approaches: the isobar model, the $K$-matrix model and a quasi-model-independent (QMI) binned approach. In the isobar model, the $S$-wave amplitude was presented by LHCb as a coherent sum of the $\sigma$ (or $f_0(500)$) meson contribution and a $\pi\pi\leftrightarrow K\ov K$ rescattering amplitude in the mass range $1.0<m_{\pi^+\pi^-}<1.5$ GeV. The fit fraction of the $S$-wave is about 25\% and predominated by the $\sigma$ resonance.

A clear \CP asymmetry was seen in the $B^-\to \pi^+\pi^-\pi^-$ decay in the following places: (i) the $S$-wave amplitude at values of $m_{\pi^+\pi^-}$ below the mass of the $\rho(770)$ resonance. In the isobar model, the $S$-wave amplitude is predominated by the $\sigma$ meson. Hence,
a significant \CP violation of 15\%  in $B^-\to\sigma\pi^-$ is implied in this model.  (ii) the $f_2(1270)$ component with  a \CP violation of 40\% exhibited, and (iii) the interference between $S$- and $P$-waves which is clearly visible in Fig. 12 of \cite{Aaij:3pi_2} where the data are split according to the sign of $\cos\theta$ with $\theta$ being the angle between the momenta of the two same-sign pions measured in the rest frame of the dipion system.
The significance of \CP violation in the interference between $S$- and $P$-waves exceeds $25\sigma$ in all the $S$-wave models.

On the contrary,
\CP asymmetry for the dominant quasi-two-body decay mode $B^-\to\rho^0\pi^-$ was found by the LHCb to  be consistent with zero in all three $S$-wave approaches (see Table \ref{tab:CPdata}),  which was already noticed by the LHCb previously in 2014 \cite{LHCb:2014}. \footnote{There was a measurement of $\acp(\rho^0\pi^-)$ by BaBar with the result $0.18\pm0.07^{+0.05}_{-0.15}$ from the Dalitz plot analysis of $B^-\to \pi^+\pi^-\pi^-$ \cite{BaBarpipipi}.}
Indeed, if this quasi-two-body \CP asymmetry is nonzero, it will destroy the aforementioned interference pattern between $S$- and $P$-waves. However, the existing theoretical
predictions based on QCD factorization (QCDF) \cite{CC:Bud,Sun:2014tfa}, perturbative QCD (pQCD) \cite{LiYa:2016}, soft-collinear effective theory (SCET) \cite{Wang:2008rk}, topological diagram approach (TDA) \cite{Cheng:TDA} and factorization-assisted topological-amplitude (FAT) approach \cite{Zhou:2016jkv} all lead to a negative \CP asymmetry for $B^-\to \rho^0\pi^-$, ranging from $-7\%$ to $-45\%$ (see Table \ref{tab:rhopiCP}).

\begin{table}[b]
\caption{Measured \CP asymmetries by the LHCb in the quasi-two-body decay $B^-\to\rho^0(770)\pi^-$ for each approach \cite{Aaij:3pi_1,Aaij:3pi_2}.  }
\label{tab:CPdata}
\begin{ruledtabular}
\begin{tabular}{ c c c c}
  & isobar & $K$-matrix & QMI  \\
\hline
$ \rho(770)^0$ & $0.7\pm1.1\pm0.6\pm1.5$ & $4.2\pm1.5\pm2.6\pm5.8$ & $4.4\pm1.7\pm2.3\pm1.6$ \\
\end{tabular}
\end{ruledtabular}
\end{table}

\begin{table}[t]
\caption{Theoretical predictions of \CP violation (in \%) for the $B^-\to \rho^0\pi^-$ decay in various approaches.}
\label{tab:rhopiCP}
\footnotesize{
\begin{ruledtabular}
\begin{tabular}{ c c c c c c}
 QCDF \cite{CC:Bud} & QCDF \cite{Sun:2014tfa} & pQCD \cite{LiYa:2016} & SCET \cite{Wang:2008rk} & TDA \cite{Cheng:TDA} & FAT \cite{Zhou:2016jkv} \\
\hline
$-9.8^{+3.4+11.4}_{-2.6-10.2}$~~ & ~~$-6.7^{+0.2+3.2}_{-0.2-3.7}$~~ & ~~$-27.5^{+2.3+0.9}_{-3.1-1.0}\pm1.4\pm0.9$~~  & ~~$-19.2^{+15.5+1.7}_{-13.4-1.9}$~~ & ~~$-23.9\pm 8.4$~~ & $-45\pm4$\\
\end{tabular}
\end{ruledtabular}}
\end{table}

The purpose of this work is twofold. First, we would like to resolve the long-standing puzzle in regard to the \CP asymmetry in  $B^-\to\rho^0\pi^-$. Second, we will present a study of $B^-\to\sigma\pi^-$ in QCDF.
For \CP violation in $B^-\to f_2(1270)\pi^-$, it has been studied in \cite{Cheng:tensor,Zou:2012td,Li:2018lbd} before the LHCb experiment.
As for \CP asymmetry induced by interference, we will give a detailed study elsewhere.

\section{$B^\pm\to \rho^0\pi^\pm$ decays}

As stressed in the Introduction, we are concerned about the discrepancy between theory and experiment in regard to \CP asymmetry in the tree-dominated mode  $B^-\to\rho^0\pi^-$.
It has been argued in \cite{Bediaga:2016} that in $B\to PV$ decays with $m_V<1$ GeV, \CP asymmetry induced from a short-distance mechanism is suppressed by the $C\!PT$ constraint.
Normally, $C\!PT$ theorem implies the same lifetimes for both particle and antiparticle.
When partial widths are summed over, the total width of the particle and its antiparticle should be the same. Final-state interactions are responsible for  distributing the \CP asymmetry among the different conjugate decay channels. In the three-body $B$ decays, the ``2+1" approximation is usually assumed so that the resonances produced in heavy meson decays do not interact with the third particle. In $B\to PV$ decays with $m_V<1$ GeV, for example, $V$=$\,\rho(770)$ or $K^*(892)$, there do not exist other states below the $K\ov K$ threshold which can be connected to $\pi\pi$ or $\pi K$ rescattering through final-state interactions. As stressed in \cite{Bediaga:2016}, the absence of final-state interactions is a hadronic constraint and therefore, the impossibility to observe \CP asymmetry in those processes is independent from the relative short-distance contribution from tree and penguin diagrams. As elucidated in \cite{Bediaga:2016}, there are three other possibilities that can produce \CP violation, for example, a three-body rescattering including the third particle.

If we take this argument seriously to explain the approximately vanishing \CP asymmetry in $B^+\to \rho^0\pi^+$, it will be at odd with the \CP violation seen in other $PV$ modeds \cite{HFLAV}: $\A_{C\!P}(B^+\to\rho^0 K^+)=0.37\pm0.11$, $\A_{C\!P}(B^0\to K^{*0}\eta)=0.19\pm0.05$ and  $\A_{C\!P}(B^0\to K^{*+}\pi^-)=-0.271\pm0.044$, especially  \CP violation in the last mode was first observed by the LHCb \cite{LHCb:Kstpi}. In general, the agreement between theory and experiment for these three modes is good (see e.g. \cite{Sun:2014tfa,CC:Bud}). Therefore, it seems to us that the smallness of $\A_{C\!P}(B^+\to\rho^0 \pi^+)$ probably has nothing to do with the $C\!PT$ constraint.

In QCDF, the decay amplitude of $B^-\to\rho^0\pi^-$ is given by \cite{BN}
\be \label{eq:Amprhopi}
A(B^-\to\rho^0\pi^-) &=& {1\over\sqrt{2}}\left[\delta_{pu}(a_2-\beta_2)-a_4^p-r_\chi^\rho a_6^p+{3\over 2}(a_7^p+a_9^p)+{1\over 2}(a_{10}^p+r_\chi^\rho a_8^p)-\beta_3^p-\beta^p_{\rm 3,EW}\right]_{\pi\rho} \non \\
&\times & X^{(B^-\pi,\rho)}
+ {1\over\sqrt{2}}\left[\delta_{pu}a_1+a_4^p-r_\chi^\pi a_6^p+a_{10}^p-r_\chi^\pi a_8^p+\beta_3^p+\beta^p_{\rm 3,EW}\right]_{\rho\pi}X^{(B^-\rho,\pi)}, \non \\
\en
where the chiral factors $r_\chi^{\pi,\rho}$ are given by
\be \label{eq:rchi}
 r_\chi^\pi(\mu)={2m_\pi^2\over m_b(\mu)(m_u+m_d)(\mu)},  \qquad r_\chi^\rho(\mu) = \frac{2m_\rho}{m_b(\mu)}\,\frac{f_\rho^\perp(\mu)}{f_\rho} \,,
\en
and the factorizable matrix elements read
\be
X^{(B^-\pi,\rho)} =2f_\rho m_B p_c F_1^{B\pi}(m_\rho^2), \qquad
X^{(B^-\rho,\pi)} =2f_\pi m_B p_c A_0^{B\rho}(m_\pi^2),
\en
with $p_c$ being the c.m. momentum. Here we have followed \cite{BSW} for the definition of form factors.
In Eq. (\ref{eq:Amprhopi}),
the order of the arguments of the $a_i^p(M_1M_2)$ and
$\beta_i(M_1M_2)$ coefficients is dictated by the subscript $M_1M_2$.

The flavor operators $a_i^{p}$ are basically the Wilson coefficients
in conjunction with short-distance nonfactorizable corrections such
as vertex corrections and hard spectator interactions. In general,
they have the expressions \cite{BBNS,BN}
 \be \label{eq:ai}
  a_i^{p}(M_1M_2) =
 \left(c_i+{c_{i\pm1}\over N_c}\right)N_i(M_2)
  + {c_{i\pm1}\over N_c}\,{C_F\alpha_s\over
 4\pi}\Big[V_i(M_2)+{4\pi^2\over N_c}H_i(M_1M_2)\Big]+P_i^{p}(M_2),
 \en
where $i=1,\cdots,10$,  the upper (lower) signs apply when $i$ is
odd (even), $c_i$ are the Wilson coefficients,
$C_F=(N_c^2-1)/(2N_c)$ with $N_c=3$, $M_2$ is the emitted meson
and $M_1$ shares the same spectator quark with the $B$ meson. The
quantities $V_i^h(M_2)$ account for vertex corrections,
$H_i^h(M_1M_2)$ for hard spectator interactions with a hard gluon
exchange between the emitted meson and the spectator quark of the
$B$ meson and $P_i(M_2)$ for penguin contractions.

In the $m_b\to\infty$ limit, the decay amplitudes of charmless two-body decays of $B$ mesons  are factorizable and can be described  in terms of decay constants and form factors. However, it is well known that the short-distance contribution to $a_4^{c,u}+r_\chi^P a_6^{c,u}$ will yield
\CP asymmetries for $\bar B^0\to K^-\pi^+, K^{*-}\pi^+, \pi^+\pi^-$, $B^-\to K^-\rho^0$ and $\bar B_s\to K^+\pi^-$,$\cdots$ etc.,  which are wrong in signs when confronted with experiment \cite{Cheng:2009,CC:Bud}. Beyond the heavy quark limit, it is thus necessary to introduce $1/m_b$ power corrections. In QCDF, power corrections to the penguin amplitudes are described by the penguin annihilation characterized by the parameters $\beta_{2,3}^p$ and $\beta^p_{\rm 3,EW}$ given in Eq. (\ref{eq:Amprhopi}). Penguin annihilation is also responsible for the rate deficit problems with penguin-dominated modes encountered in the heavy quark limit.

As pointed out in \cite{Cheng:2009,CC:Bud}, while the signs of \CP asymmetries in aforementioned modes are flipped to the right ones in the presence of power corrections from penguin annihilation,
the signs of $\acp$ in $B^-\to K^-\pi^0,~K^-\eta,~\pi^-\eta$ and $\bar B^0\to\pi^0\pi^0,~\bar K^{*0}\eta$ will also get reversed in such a way that they disagree with experiment. This \CP puzzle is resolved by invoking power corrections to the color-suppressed tree topology
as all the above-mentioned five modes receive contributions from $a_2$ \cite{Cheng:2009,CC:Bud}. An inspection of Eq. (\ref{eq:ai}) reveals that hard spectator contributions to $a_i$ are usually very small except for $a_2$ and $a_{10}$ as $c_1\sim {\cal O}(1)$ and $c_9\sim {\cal O}(-1.3)$ in units of $\alpha_{em}$. Explicitly,
\be \label{eq:a2}
a_2(M_1M_2) = c_2+{c_1\over N_c} + {c_1\over N_c}\,{C_F\alpha_s\over
 4\pi}\Big[V_2(M_2)+{4\pi^2\over N_c}H_2(M_1M_2)\Big],
\en
where the hard spectator term $H_2(M_1 M_2)$ reads
\begin{eqnarray}\label{eq:hardspec}
  H_2(M_1 M_2)= {if_B f_{M_1} f_{M_2} \over X^{(\overline{B} M_1,
  M_2)}}\,{m_B\over\lambda_B} \int^1_0 d x d y \,
 \Bigg( \frac{\Phi_{M_1}(x) \Phi_{M_2}(y)}{\bar x\bar y} + r_\chi^{M_1}
  \frac{\Phi_{m_1} (x) \Phi_{M_2}(y)}{\bar x y}\Bigg),
 \hspace{0.5cm}
 \end{eqnarray}
with $\bar x=1-x$. Subleading $1/m_b$ power corrections arise from the twist-3 amplitude $\Phi_m$. As shown in detail in \cite{Cheng:2009,CC:Bud}, power corrections to $a_2$ not only resolve the aforementioned \CP puzzles (including the so-called $\pi K$ puzzle) but also account for the  observed rates of $B^0\to \pi^0\pi^0$ and $\rho^0\pi^0$.

In the QCD factorization approach, power corrections often involve endpoint divergences. We shall follow \cite{BBNS} to model the endpoint divergence $X\equiv\int^1_0 dx/\bar x$ in the penguin annihilation and hard spectator
scattering diagrams as
\be \label{eq:XA}
 X_A^{i,f} = \ln\left({m_B\over \Lambda_h}\right)(1+\rho_A^{i,f} e^{i\phi_A^{i,f}}), \qquad
 X_H = \ln\left({m_B\over \Lambda_h}\right)(1+\rho_H e^{i\phi_H}),
\en
with $\Lambda_h$ being a typical hadronic scale of 0.5 GeV,
where the superscripts `$i$' and `$f$' refer to gluon emission from the initial and final-state quarks, respectively. In principle, one can also add the superscripts `$V\!P$' and `$PV$' to distinguish penguin annihilation effects in $B\to V\!P$ and $B\to PV$ decays \cite{BN}:
\be
&& A_1^i\approx -A_2^i\approx 6\pi\alpha_s\left[3\left(X_A^{VP,i}-4+{\pi^2\over 3}\right)+r_\chi^V r_\chi^P\Big((X_A^{VP,i})^2-2X_A^{VP,i}\Big)\right], \non \\
&& A_3^i\approx  6\pi\alpha_s\left[-3r_\chi^V\left((X_A^{VP,i})^2-2X_A^{VP,i}+4-{\pi^2\over 3}\right)+r_\chi^P \left((X_A^{VP,i})^2-2X_A^{VP,i}+{\pi^2\over 3}\right)\right], \non \\
&& A_3^f\approx  6\pi\alpha_s\left[3r_\chi^V(2X_A^{VP,f}-1)(2-X_A^{VP,f})-r_\chi^P \Big(2(X_A^{VP,f})^2-X_A^{VP,f}\Big)\right],
\en
for $M_1M_2=V\!P$  and
\be
&& A_1^i\approx -A_2^i\approx 6\pi\alpha_s\left[3\left(X_A^{PV,i}-4+{\pi^2\over 3}\right)+r_\chi^V r_\chi^P\Big((X_A^{PV,i})^2-2X_A^{PV,i}\Big)\right], \non \\
&& A_3^i\approx  6\pi\alpha_s\left[-3r_\chi^P\left((X_A^{PV,i})^2-2X_A^{PV,i}+4-{\pi^2\over 3}\right)+r_\chi^V \left((X_A^{PV,i})^2-2X_A^{PV,i}+{\pi^2\over 3}\right)\right], \non \\
&& A_3^f\approx  6\pi\alpha_s\left[-3r_\chi^P(2X_A^{PV,f}-1)(2-X_A^{PV,f})+r_\chi^V \Big(2(X_A^{PV,f})^2-X_A^{PV,f}\Big)\right],
\en
for $M_1M_2=PV$. Nevertheless,
for simplicity we shall assume that the parameters $X_A^{V\!P}$ and $X_A^{PV}$ are the same. So we shall drop the superscripts $V\!P$ and $PV$ hereafter.

Initially, it was expected that $\rho_A^i=\rho_A^f\sim 1$ and $\phi_A^i=\phi_A^f$. The two unknown parameters $\rho_A$ and $\phi_A$ were fitted to the data of $B\to PP, V\!P, PV$ and $VV$ decays. The  values of $\rho_A$ and $\phi_A$ are given, for example, in Table III of \cite{CC:Bud}, where the results are very similar to the so-called ``S4 scenario" presented in \cite{BN}. Now a surprise came from the measurement of the pure annihilation process $B_s^0\to\pi^+\pi^-$ by the CDF \cite{CDF:Bspipi} and LHCb \cite{LHCb:Bspipi}.  The world average $\B(B_s\to\pi^+\pi^-)=(0.671\pm0.083)\times 10^{-6}$ \cite{HFLAV} is much higher than the QCDF prediction of $(0.26^{+0.00+0.10}_{-0.00-0.09})\times 10^{-6}$ \cite{Cheng:Bs}.
Since this mode proceeds through the penguin-annihilation amplitudes $A_1^i$ and $A_2^i$, it is natural to expect that $\rho_A^i\neq \rho_A^f$ and that $\rho_A^i\sim 3$ is needed to accommodate the data \cite{Zhu:2011mm,Wang:2013fya}. \footnote{At first sight, the new measurement of another pure annihilation process $\B(B^0\to K^+K^-)=(7.80\pm1.27\pm0.84)\times 10^{-8}$ by the LHCb \cite{LHCb:K+K-} seems to be at odd with a large $\rho_A^i$ in the $P\!P$ sector. As can be seen from Fig. 3 in \cite{Chang:2014rla} for the dependence of $\B(B^0\to K^+K^-)$ on  $(\rho_A^{i},\phi_A^{i})$, a large $\rho_A^i$ is still allowed so long as $\phi_A^i$ is not in the region of $[-100^\circ,100^\circ]$. The constraint on the phase $\phi_A^i$ arises mainly from \CP violation in $B\to \pi K$ decays. It follows that $\phi_A^i\sim[-140^\circ,-60^\circ]$ with a large $\rho_A^i$ is favored by the data of \CP asymmetries. Putting all together, a large $\rho_A^i$ with $\phi_A^i\sim [-140^\circ,-100^\circ]$ is still favored by the data even when the new measurement of $B^0\to K^+K^-$ is take into account \cite{QChang}. }
That is, the parameters $X_A^i$ and $X_A^f$ should be treated separately. A large $\rho_A^i$ is also a good news for the hard spectator interactions because $\rho_H>3$ together a large phase $\phi_H$ are required to solve the \CP puzzle together with the rate deficit issue of $B^0\to \pi^0\pi^0$ and $\rho^0\pi^0$. Hence,  it is pertinent to set $\rho_H=\rho_A^i$ and $\phi_H=\phi_A^i$ to the first order approximation.

For $B\to PV$ decays, when $(\rho_H,\phi_H)$ and $(\rho_A^{i,f},\phi_A^{i,f})$ are treated as free parameters, it was found in \cite{Sun:2014tfa} that the allowed regions of $(\rho_A^{f},\phi_A^{f})$ are small and tight, while those of $(\rho_A^{i},\phi_A^{i})$ are big and loose. Moreover, the allowed $(\rho_H,\phi_H)$ regions are significantly separated from those of $(\rho_A^{f},\phi_A^{f})$ and overlap partly with the regions of $(\rho_A^i,\phi_A^i)$. When ($\rho_H, \phi_H)$ are set to ($\rho_A^i, \phi_A^i)$ as a first order approximation,
a fit of  the four parameters $(\rho_A^{i,f},\phi_A^{i,f})$ to the $B\to PV$ data yields \cite{Sun:2014tfa}
\be
(\rho_A^i,\rho_A^f)_{_{PV}}=(2.87^{+0.66}_{-1.95}, 0.91^{+0.12}_{-0.13}), \qquad
(\phi_A^i,\phi_A^f)_{_{PV}}=(-145^{+14}_{-21}, -37^{+10}_{-9})^\circ,
\en
where the allowed regions of ($\rho_A^i, \phi_A^i)$ shrink considerably. For comparison, they
are close to the solutions obtained in the $P\!P$ sector \cite{Chang:2014yma}
\be
(\rho_A^i,\rho_A^f)_{_{P\!P}}=(2.98^{+1.12}_{-0.86}, 1.18^{+0.20}_{-0.23}), \qquad
(\phi_A^i,\phi_A^f)_{_{P\!P}}=(-105^{+34}_{-24}, -40^{+11}_{-8})^\circ.
\en
In this work, we shall follow \cite{Chang:2015wba} to take
\be \label{eq:rhoA}
(\rho_A^i,\rho_A^f)_{_{PV}}=(3.08, 0.83), \qquad
(\phi_A^i,\phi_A^f)_{_{PV}}=(-145^\circ, -36^\circ),
\en
for calculations.

\begin{table}[t]
\caption{The branching fraction and \CP asymmetry of $B^-\to \rho^0\pi^-$ within the QCDF approach. Experimental data are taken from \cite{HFLAV}. The theoretical errors correspond to the uncertainties due to the variation of (i) Gegenbauer moments, decay constants,  form factors, the strange quark mass, and (ii) $\rho_{A,H}$, $\phi_{A,H}$, respectively. In (ii) we assign an error of $\pm0.4$ to $\rho$ and $\pm 4^\circ$ to $\phi$.}
\label{tab:rhopi_theory}
\begin{center}
\begin{tabular}{ l c l} \hline \hline
 $\B(10^{-6})$ & ~~$\A_{C\!P}(\%)$~~ &  Comments \\
\hline
 $8.3^{+1.2}_{-1.3}$ & ~~~~~$0.7\pm1.9$~~~~~ & Expt \\
 \hline
 $8.9^{+2.0+0.0}_{-1.0-0.0}$ & ~~$6.3^{+0.5+0.0}_{-0.8-0.0}$ & (1) Heavy~quark~limit \\
 $9.3^{+1.8+0.3}_{-1.0-0.3}$ & $-13.0^{+1.0+3.5}_{-0.8-3.8}$ & (2) $\rho_{H}=0$ and $\phi_{H}=0$~ with~$\rho_A$ and $\phi_A$ given by Eq. (\ref{eq:rhoA})  \\
 $6.7^{+0.6+0.2}_{-0.4-0.2}$ & $-4.8^{+4.3+3.8}_{-2.4-4.1}$ & (3) $\rho_H=3.08,~\phi_H=-145^\circ$, $\rho_A$ and $\phi_A$ given by Eq. (\ref{eq:rhoA})   \\
 $8.4^{+1.6+0.2}_{-0.8-0.2}$ & $-0.7^{+4.3+3.2}_{-2.8-3.5}$ & (4) $\rho_H=3.15$, ~$\phi_H=-113^\circ$, $\rho_A$ and $\phi_A$ given by Eq. (\ref{eq:rhoA}) \\
 $6.4^{+0.6+0.2}_{-0.4-0.2}$ & $14.4^{+2.2+1.1}_{-1.3-1.0}$ &  (5) $\rho_H=3.08,~\phi_H=-145^\circ$, $\rho_A^{i,f}=0$, $\phi_A^{i,f}=0$ \\
 $8.1^{+1.7+0.2}_{-0.8-0.2}$ & $15.2^{+1.3+1.0}_{-1.1-1.0}$ &  (6) $\rho_H=3.15,~\phi_H=-113^\circ$, $\rho_A^{i,f}=0$, $\phi_A^{i,f}=0$ \\
\hline \hline
\end{tabular}
\end{center}
\end{table}

We are now ready to compute the branching fraction and \CP asymmetry for $B^-\to \rho^0\pi^-$. In the heavy quark limit, its \CP asymmetry is positive with a magnitude of order 0.06\,. We then turn on power corrections induced from penguin annihilation. It is clear that the sign of $\acp(\rho^0\pi^-)$ is flipped and in the meantime its magnitude is enhanced. We next switch on $1/m_b$ corrections from hard spectator interactions. Under the simplification with $\rho_H=\rho_A^i$ and $\phi_H=\phi_A^i$, we will have $\B(\rho^0\pi^-)\approx 6.7\times 10^{-6}$ and $\acp(\rho^0\pi^-)\approx -0.05$\,.  However, the resultant  branching fraction is too small by 20\% when compared with experiment. This implies that the realistic values of $\rho_H$ and $\phi_H$ should have some deviation from $\rho_A^i$ and $\phi_A^i$, respectively.
Indeed, we find that the data can be accommodated by having $\rho_H=3.15$ and $\phi_H=-113^\circ$, for instance, shown in case (4) of Table \ref{tab:rhopi_theory}. To see the effect of hard spectator interactions alone, we turn off $\rho_A$ and $\phi_A$. It is evident  that $\acp(\rho^0\pi^-)$ will be enhanced from ${\cal O}(6)$ to ${\cal O}(15)$ in the presence of hard spectator effects.
If the heavy quark limit of $\acp(\rho^0\pi^-)$ is considered as a benchmark, hard spectator interactions will push it up  further, whereas penguin annihilation will pull it to the opposite direction. Therefore, the nearly vanishing  $\acp(\rho^0\pi^-)$ arises from two destructive  $1/m_b$ power corrections.

What about the previous QCDF predictions given in Table \ref{tab:rhopiCP}? The results of
$\acp(\rho^0\pi^-)\approx -0.098$ and $\B(\rho^0\pi^-)\approx 8.7\times 10^{-6}$ given in \cite{CC:Bud} were obtained using $\rho_A\sim 1$ and
$\phi_A^{VP}=-70^\circ$ and $\phi_A^{PV}=-30^\circ$, while the power correction to $a_2$ was parameterized as $(1+0.8e^{-i80^\circ})$.
The QCDF predictions  $\acp(\rho^0\pi^-)\approx -0.067$ and $\B(\rho^0\pi^-)\approx 6.8\times 10^{-6}$ given in \cite{Sun:2014tfa} are very similar to case (3) in Table \ref{tab:rhopi_theory}.
As noticed in passing, one needs to adjust $\rho_H$ and $\phi_H$ slightly to render both the branching fraction and \CP asymmetry in agreement with the data.

\section{$B^\pm\to \sigma\pi^\pm$ decays}

Charmless hadronic $B$ decays to scalar mesons have been studied  in the approach of QCD factorization \cite{Cheng:scalar,Cheng:2007st,CCY:SP}. For completeness, we shall present a study of $B^-\to \sigma/f_0(500)\pi^-$.
Its decay amplitude  has a similar expression as $B^-\to f_0(980)\pi^-$ in Eq. (A1) of  \cite{Cheng:scalar}:
\be \label{eq:sigmapi}
A(B^- \to \sigma \pi^- ) &=&
 \frac{G_F}{\sqrt{2}}\sum_{p=u,c}\lambda_p^{(d)}
 \Bigg\{ \left[a_1 \delta_{pu}+a^p_4+a_{10}^p-(a^p_6+a^p_8) r_\chi^\pi \right]_{\sigma\pi} X^{(B\sigma,\pi)} \non \\
 &+&
 \left[a_2\delta_{pu} +2(a_3^p+a_5^p)+{1\over 2}(a_7^p+a_9^p)+a_4^p-{1\over 2}a_{10}^p-(a_6^p-{1\over 2}a_8^p)\bar r^\sigma_\chi\right]_{\pi\sigma}\ov X^{(B\pi,\sigma^u)}\non \\
 &-& f_Bf_\pi\bar f_{\sigma}^u\bigg[\delta_{pu}b_2(\pi\sigma)+ b_3(\pi\sigma)
 + b_{\rm 3,EW}(\pi\sigma) +(\pi\sigma\to \sigma\pi) \bigg] \Bigg\},
\en
where the factorizable matrix elements read
\be
X^{(B\sigma,\pi)}=-f_\pi F_0^{B\sigma^u}(m_\pi^2)(m_B^2-m_\sigma^2), \qquad
\ov X^{(B\pi,\sigma)}=\bar f_\sigma^u F_0^{B\pi}(m_\sigma^2)(m_B^2-m_\pi^2),
\en
with $\bar r_\chi^{\sigma}(\mu)=2m_{\sigma}/m_b(\mu)$ and  $\lambda_p^{(d)}=V_{pb}V_{pd}^*$.
The superscript $u$ in the scalar decay constant $\bar f_\sigma^u$ and the form factor $F^{B\sigma^u}$ refers to the $u$ quark component of the $\sigma$.

\begin{table}[t]
\caption{Numerical values of the flavor operators $a_i^p[M_1M_2]$ for $M_1M_2=\sigma\pi$ and $\pi\sigma$ at the scale $\mu=m_b(m_b)$. Penguin annihilation characterized by the parameter $\beta^p$ defined by Eq. (\ref{eq:beta}) is also shown.}
\label{tab:ai}
\begin{center}
\begin{tabular}{ l c c | l c c} \hline \hline
 $a_i$ & ~~$\sigma\pi$~~ & ~~~$\pi\sigma$~~~ & ~~$a_i$~~  & ~~$\sigma\pi$~~ & ~~$\pi\sigma$ \\
\hline
 $a_1$ & ~~~$0.95+0.014i$~~~ & ~~$0.015-0.004i$~~ & ~~$a_7$ & $(-1.8+0.3i)10^{-4}$ & $(-4.2+1.0i)10^{-5}$ \\
 $a_2$ & $0.33-0.080i$ & $-0.056+0.024i$ & ~~$a_8^u$ & $(4.8-1.0i)10^{-4}$ & $(4.8-1.0i)10^{-4}$  \\
 $a_3$ & $-0.009+0.003i$ & $0.0026-0.0008i$ & ~~$a_8^c$ & $(4.6-0.5i)10^{-4}$  &  $(4.6-0.5i)10^{-4}$\\
 $a_4^u$ & $-0.022-0.015i$ & $0.062-0.013i$ & ~~$a_9$  & $(-8.6-0.1i)10^{-3}$ & $(-1.3+0.4i)10^{-4}$\\
 $a_4^c$ & $-0.027-0.006i$ & $-0.012-0.007i$ & ~~$a_{10}^u$ & $(-2.6+0.6i)10^{-3}$  & $(8.7-3.1i)10^{-4}$ \\
 $a_5$ & $0.0158-0.003i$ & $0.0035-0.0009i$ & ~~$a_{10}^c$ & $(-2.6+0.7i)10^{-3}$ & $(4.6-2.8i)10^{-4}$ \\
 $a_6^u$ & $-0.042-0.014i$ & $-0.042-0.014i$  & ~~$\beta^p$ & $(-1.6-0.1i)10^{-4}$ &
 $(3.6+3.2i)10^{-5}$ \\
 $a_6^c$ & $-0.045-0.005i$ & $-0.045-0.005i$ & \\
\hline \hline
\end{tabular}
\end{center}
\end{table}

It is known that the neutral scalar meson $\sigma$ cannot be produced via the vector current. If $\sigma$ is a 2-quark bound state with the flavor wave function $(\bar uu+\bar dd)/\sqrt{2}$, its scale-dependent scalar decay constant can be defined as
\be
\la\sigma |\bar uu|0\ra=m_\sigma \bar f_\sigma^u. 
\en
For simplicity, we will not consider the mixing of $\sigma$ and $f_0(980)$ and hence the strange quark effect in Eq. (\ref{eq:sigmapi}).
In this work we shall assume that $\sigma$ has a similar decay constant and light-cone distribution amplitude (LCDA) as $f_0(980)$. Explicitly, we  take $\bar f_\sigma^u=350$ MeV at $\mu=1$ GeV and $F_0^{B\sigma^u}(0)=0.25$, where the Clebsch-Gordon coefficient $1/\sqrt{2}$ is included in $\bar f_\sigma^u$ and $F_0^{B\sigma^u}$. Vertex corrections, hard spectator interactions and weak annihilation for $B\to SP$ and $B\to SV$ have been worked out in \cite{Cheng:scalar,Cheng:2007st,CCY:SP}. Since the twist-2 LCDA of the $\sigma$ meson is dominated by the odd Gengenabauer moments, which vanish for the $\pi$ mesons, it follows that the flavor operators $a_i^p(\pi\sigma)$ and $a_i^p(\sigma\pi)$ can be very different numerically except for $a_{6,8}^p$ (see Table \ref{tab:ai}). For example, $a_1(\sigma\pi)\approx 1\gg a_1(\pi\sigma)$. It appears that $a_i^p(\sigma\pi)$ look like the normal ones, but  not $a_i^p(\pi\sigma)$.
Effects of penguin annihilation defined by
\be  \label{eq:beta}
\beta^p(M_1M_2)=-f_Bf_\pi\bar f_{\sigma}^u\big[\delta_{pu}b_2+ b_3
 + b_{\rm 3,EW}\big]_{M_1M_2}
\en
are also shown in Table \ref{tab:ai}.

Using the input parameters given in \cite{Cheng:scalar} except for the Wolfenstin parameters updated with $A=0.8235$, $\lambda=0.224837$, $\bar \rho=0.1569$ and $\bar
\eta=0.3499$ \cite{CKMfitter}, we obtain
\be
\B(B^-\to\sigma\pi^-)=(5.38^{+0.19+1.34+0.94}_{-0.18-1.20-0.90})\times 10^{-6}, \qquad
\A_{C\!P}(B^-\to\sigma\pi^-)=(15.95^{+0.29+0.08+18.88}_{-0.28-0.06-21.88})\%. \non \\
\en
Theoretical uncertainties come from (i) the Gegenbauer moments $B_{1,3}$, the scalar meson decay constants, (ii) the heavy-to-light form factors and
the strange quark mass, and (iii) the power corrections due to weak annihilation and hard spectator interactions, respectively.
The calculated \CP asymmetry agrees well with the LHCb measurement \cite{Aaij:3pi_1,Aaij:3pi_2}
\be
\A_{C\!P}(B^-\to\sigma\pi^-)=(16.0\pm1.7\pm2.2)\%.
\en

From the fit fraction $(25.2\pm0.5\pm5.0)\%$ of the $\sigma$ component in $B^-\to\pi^+\pi^-\pi^-$ decay analyzed in the isobar model \cite{Aaij:3pi_1,Aaij:3pi_2} and the total branching fraction $(15.2\pm1.4)\times 10^{-6}$ measured by BaBar \cite{BaBarpipipi}, we obtain
\be
\B(B^-\to\sigma\pi^-\to \pi^+\pi^-\pi^-)_{\rm expt}=(3.83\pm0.76)\times 10^{-6}.
\en
To compute the decay rate of $B^-\to\sigma\pi^-\to \pi^+\pi^-\pi^-$ it is necessary to take into account the resonance shape of the $\sigma$, for example, the standard Breit-Wigner function.  If $\sigma$ were very narrow, one would have the narrow width approximation
\be
\B(B^-\to\sigma\pi^-\to \pi^+\pi^-\pi^-)=\B(B^-\to\sigma\pi^-)\B(\sigma\to \pi^+\pi^-).
\en
Since $\B(\sigma\to \pi^+\pi^-)\approx 2/3$, it appears that the above relation is empirically working. However, as $\sigma$ is very broad, its finite width effect could be very important \cite{Qi:2018lxy}.

\section{Conclusions}
The decay amplitudes of $B^+\to \pi^+\pi^-\pi^+$ in the Dalitz plot have been analyzed by the LHCb using three different approaches for the $S$-wave component. It was found that the mode with $\sigma$ (or $f_0(500)$) exhibited a \CP asymmetry of 15\% in the isobar model, whereas the $f_2(1270)$ mode had a 40\% asymmetry. In contrast,
\CP asymmetry for the dominant quasi-two-body decay $B^-\to\rho^0\pi^-$ was found to  be consistent with zero in all three approaches, while all the existing theoretical
predictions  lead to a negative \CP asymmetry ranging from $-7\%$ to $-45\%$. We show that the nearly vanishing \CP violation in $B^-\to\rho^0\pi^-$ is understandable in QCDF.  The $1/m_b$ power corrections penguin annihilation and hard spectator interactions contribute destructively to $\acp(B^-\to\rho^0\pi^-)$ to render it consistent with zero. The branching fraction and \CP asymmetry in $B^-\to\sigma/f_0(500)\pi^-$ are investigated in QCDF with results in agreement with experiment.

\vskip 2.0cm \acknowledgments

We are very grateful to Qin Chang for helpful discussions.
This research was supported in part by the Ministry of Science and Technology of R.O.C. under Grant No. 107-2119-M-001-034.



\begin{thebibliography}{99}
\newcommand{\bi}{\bibitem}

\bibitem{LHCb:Kppippim} R. Aaij {\it et al.} [LHCb Collaboration],
  ``Measurement of CP violation in the phase space of $B^{\pm} \to K^{\pm} \pi^{+} \pi^{-}$ and $B^{\pm} \to K^{\pm} K^{+} K^{-}$ decays,''
  Phys.\  Rev.\  Lett.\  {\bf 111}, 101801 (2013)
  [arXiv:1306.1246 [hep-ex]].

\bibitem{LHCb:pippippim}
   R.~Aaij {\it et al.} [LHCb Collaboration],
  ``Measurement of CP violation in the phase space of $B^{\pm} \rightarrow K^{+} K^{-} \pi^{\pm}$ and $B^{\pm} \rightarrow \pi^{+} \pi^{-} \pi^{\pm}$ decays,''
  Phys.\ Rev.\ Lett.\  {\bf 112},  011801 (2014)
  [arXiv:1310.4740 [hep-ex]].

\bibitem{LHCb:2014}
  R.~Aaij {\it et al.} [LHCb Collaboration],
  ``Measurements of $CP$ violation in the three-body phase space of charmless $B^{\pm}$ decays,''
  Phys.\ Rev.\ D {\bf 90},  112004 (2014)
  [arXiv:1408.5373 [hep-ex]].

\bibitem{Aaij:3pi_1}
  R.~Aaij {\it et al.} [LHCb Collaboration],
  ``Observation of Several Sources of $CP$ Violation in $B^+ \to \pi^+ \pi^+ \pi^-$ Decays,''
  Phys.\ Rev.\ Lett.\  {\bf 124},  031801 (2020)
  [arXiv:1909.05211 [hep-ex]].

\bibitem{Aaij:3pi_2}
  R.~Aaij {\it et al.} [LHCb Collaboration],
  ``Amplitude analysis of the $B^+ \rightarrow \pi^+\pi^+\pi^-$ decay,''
  Phys.\ Rev.\ D {\bf 101},  012006 (2020)
  [arXiv:1909.05212 [hep-ex]].

\bibitem{BaBarpipipi}
  B.~Aubert {\it et al.}   [BaBar Collaboration],
  ``Dalitz Plot Analysis of $B^+\to \pi^+\pi^+\pi^-$ Decays,''
  Phys.\ Rev.\ D {\bf 79}, 072006 (2009)  [arXiv:0902.2051 [hep-ex]].

\bibitem{CC:Bud}
  H.~Y.~Cheng and C.~K.~Chua,
  ``Revisiting Charmless Hadronic $B_{u,d}$ Decays in QCD Factorization,''
  Phys.\ Rev.\ D {\bf 80}, 114008 (2009)
  [arXiv:0909.5229 [hep-ph]].

\bibitem{Sun:2014tfa}
J.~Sun, Q.~Chang, X.~Hu and Y.~Yang,
``Constraints on hard spectator scattering and annihilation corrections in $B_{u,d}$ ${\to}$ $PV$ decays within QCD factorization,''
Phys. Lett. B \textbf{743}, 444-450 (2015)
[arXiv:1412.2334 [hep-ph]].

\bibitem{LiYa:2016}
Y.~Li, A.~Ma, W.~Wang and Z.~Xiao,
``Quasi-two-body decays $B_{(s)}\to P\rho\to P\pi\pi$ in perturbative QCD approach,''
Phys. Rev. D \textbf{95},  056008 (2017)
[arXiv:1612.05934 [hep-ph]].

\bibitem{Wang:2008rk}
  W.~Wang, Y.~M.~Wang, D.~S.~Yang and C.~D.~Lu,
  ``Charmless Two-body $B_{(s)} \to VP$ decays In Soft-Collinear-Effective-Theory,''
  Phys.\ Rev.\ D {\bf 78}, 034011 (2008)
  [arXiv:0801.3123 [hep-ph]].

\bibitem{Cheng:TDA}
H.~Cheng, C.~Chiang and A.~Kuo,
``Updating $B\to PP,VP$ decays in the framework of flavor symmetry,''
Phys. Rev. D \textbf{91},  014011 (2015)
[arXiv:1409.5026 [hep-ph]].

\bibitem{Zhou:2016jkv}
S.~Zhou, Q.~Zhang, W.~Lyu and C.~Lü,
``Analysis of Charmless Two-body B decays in Factorization Assisted Topological Amplitude Approach,''
Eur. Phys. J. C \textbf{77},  125 (2017)
[arXiv:1608.02819 [hep-ph]].

\bibitem{Cheng:tensor}
H.~Y.~Cheng and K.~C.~Yang,
``Charmless Hadronic $B$ Decays into a Tensor Meson,''
Phys. Rev. D \textbf{83}, 034001 (2011)
[arXiv:1010.3309 [hep-ph]]

\bibitem{Zou:2012td}
Z.~T.~Zou, X.~Yu and C.~D.~Lu,
``Nonleptonic two-body charmless $B$ decays involving a tensor meson in the Perturbative QCD Approach,''
Phys. Rev. D \textbf{86}, 094015 (2012)
[arXiv:1203.4120 [hep-ph]].

\bibitem{Li:2018lbd}
Y.~Li, A.~J.~Ma, Z.~Rui, W.~F.~Wang and Z.~J.~Xiao,
``Quasi-two-body decays $B_{(s)}\to P f_2(1270)\to P\pi\pi$ in the perturbative QCD approach,''
Phys. Rev. D \textbf{98},  056019 (2018)
[arXiv:1807.02641 [hep-ph]].

\bibitem{Bediaga:2016}
  J.~H.~Alvarenga Nogueira, I.~Bediaga, T.~Frederico, P.~C.~Magalhães and J.~Molina Rodriguez,
  ``Suppressed $B \to PV $ CP asymmetry: CPT constraint,''
  Phys.\ Rev.\ D {\bf 94},  054028 (2016)
  [arXiv:1607.03939 [hep-ph]].

\bibitem{HFLAV} Y.~Amhis {\it et al.} [HFLAV Collaboration],
  ``Averages of $b$-hadron, $c$-hadron, and $\tau$-lepton properties as of summer 2016,''
  Eur.\ Phys.\ J.\ C {\bf 77},  895 (2017)
  [arXiv:1612.07233 [hep-ex]],
  and online update at https://hflav.web.cern.ch\,.

\bibitem{LHCb:Kstpi}
R.~Aaij \textit{et al.} [LHCb Collaboration],
``Amplitude analysis of the decay $\overline{B}^0 \to K_{S}^0 \pi^+ \pi^-$ and first observation of the CP asymmetry in $\overline{B}^0 \to K^{*}(892)^- \pi^+$,''
Phys. Rev. Lett. \textbf{120}, 261801 (2018)
[arXiv:1712.09320 [hep-ex]].


\bibitem{BN} M. Beneke and M. Neubert, 
``QCD factorization for $B\to PP$ and $B \to PV$ decays,''
Nucl. Phys. B \textbf{675}, 333-415 (2003)
[arXiv:hep-ph/0308039 [hep-ph]].

\bibitem{BSW} M. Wirbel, B. Stech, and M. Bauer, 
``Exclusive Semileptonic Decays of Heavy Mesons,''
Z. Phys. C \textbf{29}, 637 (1985); 
M. Bauer, B. Stech, and M. Wirbel, 
``Exclusive Nonleptonic Decays of $D$, $D_s$, and $B$ Mesons,''
Z. Phys. C \textbf{34}, 103 (1987) C {\bf 34},
103 (1987).

\bi{BBNS} M. Beneke, G. Buchalla, M. Neubert, and C.T. Sachrajda,
``QCD factorization for $B \to PP$ decays: Strong phases and CP violation in the heavy quark limit,''
Phys. Rev. Lett. \textbf{83}, 1914-1917 (1999)
[arXiv:hep-ph/9905312 [hep-ph]];
``QCD factorization for exclusive, nonleptonic B meson decays: General arguments and the case of heavy light final states,''
Nucl. Phys. B \textbf{591}, 313-418 (2000)
[arXiv:hep-ph/0006124 [hep-ph]].

\bibitem{Cheng:2009}
H.~Y.~Cheng and C.~K.~Chua,
``Resolving B-CP Puzzles in QCD Factorization,''
Phys. Rev. D \textbf{80}, 074031 (2009)
[arXiv:0908.3506 [hep-ph]].

\bibitem{CDF:Bspipi}
T.~Aaltonen \textit{et al.} [CDF Collaboration],
``Evidence for the charmless annihilation decay mode $B^0_s \to \pi^+\pi^-$,''
Phys. Rev. Lett. \textbf{108}, 211803 (2012)
[arXiv:1111.0485 [hep-ex]].


\bibitem{LHCb:Bspipi}
R.~Aaij \textit{et al.} [LHCb Collaboration],
``Observation of the annihilation decay mode $B^{0}\to K^{+}K^{-}$,''
Phys. Rev. Lett. \textbf{118},  081801 (2017)
[arXiv:1610.08288 [hep-ex]].

\bibitem{Cheng:Bs}
  H.~Y.~Cheng and C.~K.~Chua,
  ``{QCD} Factorization for Charmless Hadronic $B_s$ Decays Revisited,''
  Phys.\ Rev.\ D {\bf 80}, 114026 (2009)
  [arXiv:0910.5237 [hep-ph]].

\bibitem{Zhu:2011mm}
G.~Zhu,
``Implications of the recent measurement of pure annihilation $B_s \to \pi^+ \pi^-$ decays in QCD factorization,''
Phys. Lett. B \textbf{702}, 408-412 (2011)
[arXiv:1106.4709 [hep-ph]].

\bibitem{Wang:2013fya}
K.~Wang and G.~Zhu,
``Flavor dependence of annihilation parameters in QCD factorization,''
Phys. Rev. D \textbf{88}, 014043 (2013)
[arXiv:1304.7438 [hep-ph]].

\bibitem{LHCb:K+K-}
R.~Aaij \textit{et al.} [LHCb Collaboration],
``Observation of the annihilation decay mode $B^{0}\to K^{+}K^{-}$,''
Phys. Rev. Lett. \textbf{118}, 081801 (2017)
[arXiv:1610.08288 [hep-ex]].

\bibitem{Chang:2014rla}
Q.~Chang, J.~Sun, Y.~Yang and X.~Li,
``Spectator scattering and annihilation contributions as a solution to the $\pi K$ and $\pi \pi$ puzzles within QCD factorization approach,''
Phys. Rev. D \textbf{90}, 054019 (2014)
[arXiv:1409.1322 [hep-ph]].

\bibitem{QChang} Qin Chang, private communication.


\bibitem{Chang:2014yma}
Q.~Chang, J.~Sun, Y.~Yang and X.~Li,
``A combined fit on the annihilation corrections in $B_{u,d,s}\to PP$ decays within QCDF,''
Phys. Lett. B \textbf{740}, 56-60 (2015)
[arXiv:1409.2995 [hep-ph]].

\bibitem{Chang:2015wba}
  Q.~Chang, X.~Hu, J.~Sun and Y.~Yang,
  ``Probing spectator scattering and annihilation corrections in $B_{s} \to PV$ decays,''
  Phys.\ Rev.\ D {\bf 91}, 074026 (2015)
  [arXiv:1504.04907 [hep-ph]].


\bibitem{CCY:SP}
  H.~Y.~Cheng, C.~K.~Chua and K.~C.~Yang,
  ``Charmless hadronic $B$ decays involving scalar mesons: Implications to the
  nature of light scalar mesons,''
  Phys.\ Rev.\  D {\bf 73}, 014017 (2006)
  [arXiv:hep-ph/0508104].

\bibitem{Cheng:2007st}
H.~Y.~Cheng, C.~K.~Chua and K.~C.~Yang,
``Charmless B decays to a scalar meson and a vector meson,''
Phys. Rev. D \textbf{77}, 014034 (2008)
[arXiv:0705.3079 [hep-ph]].

\bibitem{Cheng:scalar}
  H.~Y.~Cheng, C.~K.~Chua, K.~C.~Yang and Z.~Q.~Zhang,
  ``Revisiting charmless hadronic $B$ decays to scalar mesons,''
  Phys.\ Rev.\ D {\bf 87},  114001 (2013)
  [arXiv:1303.4403 [hep-ph]].

\bibitem{CKMfitter}  J. Charles {\it et al.} [CKMfitter Group], Eur.
Phys. J. C {\bf 41}, 1 (2005) [hep-ph/0406184], updated results and plots available at: http://ckmfitter.in2p3.fr;  M. Bona {\it et
al.} [UTfit Collaboration], JHEP {\bf 0507}, 028 (2005) and updated results from
http://utfit.roma1.infn.it.

\bibitem{Qi:2018lxy}
J.~J.~Qi, Z.~Y.~Wang, X.~H.~Guo and Z.~H.~Zhang,
``Study of localized $CP$ violation in $B^-\rightarrow \pi^- \pi^+\pi^-$ and the branching ratio of $B^-\rightarrow \sigma(600)\pi^-$ in the QCD factorization approach,''
Nucl. Phys. B {\bf 948}, 114788 (2019)
[arXiv:1811.10333 [hep-ph]].  

\end{thebibliography}
\end{document}